\documentclass[12pt]{article}
\usepackage{amsfonts}
\usepackage{amssymb}
\usepackage{latexsym}
\usepackage{graphicx}
\usepackage[english]{babel}
\input epsf
\long\def\symbolfootnote[#1]#2{\begingroup%
\def\thefootnote{\fnsymbol{footnote}}\footnote[#1]{#2}\endgroup}

\begin{document}

 \renewcommand{\theequation}{\arabic{section}.\arabic{equation}}
 \newcommand{\newsection}[1]{\section{#1} \setcounter{equation}{0}}

\def\be{\begin{equation}}
\def\bea{\begin{eqnarray}}
\def\ee{\end{equation}}
\def\eea{\end{eqnarray}}

\def\Hbar{{\bar H}}
\def\Htilde{{\tilde H}}
\def\btilde{{\tilde \beta}}
\def\otilde{{\tilde \omega}}
\def\bbar{{\bar \beta}}
\def\obar{{\bar \omega}}
\def\Ftilde{{\tilde \mathcal{F}}}
\def\htilde{\tilde h}
\def\Vbar{{\bar V}}
\def\Vtilde{{\tilde V}}
\def\chibar{{\bar \chi}}
\def\chitilde{{\tilde \chi}}
\def\ebar{\hat{{\bar e}}}
\def\etilde{\hat{{\tilde e}}}
\def\btilde{\tilde\beta}
\def\otilde{\tilde\omega}

\def\abar{{\bar a}}
\def\d3bar{{\bar d}_3}
\def\F{\mathcal{F}}
\def\ehat{\hat{e}}
\def\D{\mathcal{D}}
\def\G{\mathcal{G}^+}
\def\bstar{\bar\star}
\def\field{\mathbb}
 \newcommand{\w}{\wedge}
 \renewcommand{\star}{*}
 \newcommand{\tr}{\tilde{r}}
 \newcommand{\ttheta}{\tilde{\theta}}
 \newcommand{\tphi}{\tilde{\phi}}
 \newcommand{\tpsi}{\tilde{\psi}}
\newcommand{\ta}{\tilde{a}}
\newcommand{\tb}{\tilde{b}}
\newcommand{\tsigma}{\tilde{\sigma}}
\vspace{20mm}
\begin{center} {\LARGE A class of BPS time-dependent 3-charge microstates from spectral flow }
\\
\vspace{20mm} {\bf  Jon Ford, Stefano Giusto and Ashish Saxena}\\ 
\vspace{2mm}
\symbolfootnote[0]{ {\tt jford@physics.utoronto.ca, giusto@physics.utoronto.ca, ashish@physics.utoronto.ca}} 
Department of Physics,\\ University of Toronto,\\ 
Toronto, Ontario, Canada M5S 1A7\\
\vspace{4mm}
\end{center}
\vspace{10mm}
\begin{abstract}
We construct an infinite family of asymptotically flat 3-charge solutions carrying D1, D5 and momentum charges. Generically the solutions also carry two angular momenta. The geometries describe the spectral flow of all the ground states of the D1-D5 CFT. The family is parametrized by four functions describing the embedding of a closed curve in $\field{R}^4$ and an integer $n$ labelling the spectral flow on the left sector. After giving the general prescription for spectral flowing any of the ground states, we give an explicit example of the construction. We identify the asymptotic charges of the resulting solution and show the matching with the corresponding CFT result.   
\end{abstract}

\thispagestyle{empty}
\newpage
\setcounter{page}{1}

\newsection{Introduction}

The Mathur conjecture \cite{mss, MathurRecentReview} provides a gravitational interpretation of the thermodynamical characteristics of horizons and black hole systems. The conjecture proposes that instead of a single classical black hole metric, there exists an ensemble of horizon-free geometries, each of which possesses the same conserved charges and asymptotics as the original black hole. In this model the horizon and the associated Bekenstein-Hawking entropy should arise due to a statistical averaging over the ensemble \cite{lmstat}. Results from string theory, specifically the AdS/CFT duality \cite{Maldacena,Witten}, have made it possible to explicitly find some of these geometries and to verify that they possess the necessary properties to be classified as gravitational microstates. 

So far the strongest evidence in support of the Mathur conjecture has come from the study of the 2-charge D1-D5 system, which consists of D1 and D5-branes wrapped on a five torus. Although this system does not produce a black hole with a classical horizon of finite area, it does possess degenerate states and therefore entropy. This entropy can be shown to result from a stretched 
horizon generated by quantum corrections \cite{sen,dabholkar}. In \cite{LM} Lunin and Mathur  constructed a family of regular microstate geometries by performing a series of duality operations that mapped the D1-D5 system to a system consisting of a fundamental string with winding and momentum charges. In this construction the gravity duals were parametrized in terms of a closed curve in $\field{R}^4$ that characterizes the profile of the oscillating string. It has been shown that these geometries are non-singular (with the exception of acceptable orbifold singularities) and horizon-free, and that this family of microstate geometries accounts for a finite fraction of the entropy of the 2-charge black hole \cite{lmm}. In \cite{lms} it was shown that the chiral primaries in the NS sector are related to these solutions by a spectral flow transformation. The work on the 2-charge system has since been continued in \cite{taylor1,taylor2}, in which the complete set of microstates has been found and a precise mapping rule between the CFT states and the gravity duals has been proposed. 

The 2-charge system provides substantial motivation for the Mathur proposal. Yet another non-trivial check on the conjecture is provided by considering the 3-charge system  obtained by adding momentum along the D1-branes. This D1-D5-P system is the prototypical example of a 5D BPS black hole in string theory. Its naive geometry in the gravity picture is an extremal Reissner-Nordstrom type black hole, which possesses macroscopic entropy and a regular horizon \cite{bmpv}. The D1-D5-P system has been analyzed extensively in both the gravity and the field theory pictures \cite{cy, Larsen5D}. In \cite{stromvafa} the entropy was computed microscopically by looking at the degrees of freedom in the world volume theory and exact agreement with the Bekenstein-Hawking entropy was found. Since the 3-charge system possesses all the physically relevant characteristics of a black hole (i.e. macroscopic entropy and a horizon), finding the microstates of the D1-D5-P system would represent a compelling test of validity of the Mathur conjecture. Unfortunately constructing microstates in the 3-charge case poses a significant technical challenge. The techniques used in \cite{LM} cannot be repeated because dualities cannot be used to map the D1-D5-P system to a simpler one. Some progress  has been made in understanding 3-charge systems as supertubes in M-theory \cite{supertubes,supertubes1,supertubes2,superdupertubes}. Despite the difficulty, gravity duals for some simple D1-D5-P states have been found. For example, in \cite{mss} perturbative techniques were used to add a single unit of momentum charge to a particular Ramond ground state and the resulting solution was shown to be regular. In \cite{Lunin,gms1,gms2} axially-symmetric microstates were constructed by adding momentum to the simplest members of the 2-charge D1-D5 ensemble. Again, these geometries were found to be horizon-free and to possess the necessary properties of black hole microstates proposed by the Mathur conjecture. In \cite{simon1} some non-BPS microstates were found and shown to be smooth. More general classes of 3-and 4-charge BPS solutions possessing at least one axial isometry have been found in~\cite{others}. However, the solutions mentioned above represent only a small subset of the 3-charge microstates and further work in this area is necessary to understand the microscopic description of black hole entropy. 

In this article we present a new class of time-dependent, BPS gravity duals for the D1-D5-P system, consisting of an infinite family of solutions parameterized by an integer and a closed curve in $\field{R}^4$. This class of solutions represent the spectral flow of all Ramond ground states of the D1-D5 field theory. The procedure we use to construct these solutions is summarized by the following basic steps:
\begin{itemize}
\item We start with the supergravity solutions dual to the set of Ramond  ground states of the 2-charge D1-D5 system. We then set the charges equal and express the geometry in the language of six-dimensional minimal supergravity. 
\item After taking the near-horizon limit of the 2-charge geometries we add momentum by applying a spectral flow transformation on the left sector of the CFT \cite{bal,mm}. This produces a 3-charge geometry with $AdS_3\times S^3$ asymptotics.
\item We then find an extension of the solution that produces a new asymptotically flat geometry satisfying the equations of~\cite{GMR}.
\end{itemize}
A similar procedure was used in \cite{Lunin} to construct an axially-symmetric 3-charge solution.

The layout of this paper is as follows. In \S{2} we review the generic form of solutions in  six-dimensional minimal supergravity and introduce the GMR notation \cite{GMR} and equations of motion. In \S{3} we give the 2-charge metrics that are dual to the Ramond ground states of the D1-D5 system, re-write them in the GMR notation  and reduce to six-dimensions. In \S{4} we add momentum to the system by applying a spectral flow transformation and re-diagonalize into the GMR form. In \S{5} we show how the 3-charge near-horizon solution can be extended to an asymptotically flat solution. In \S{6} we present an explicit example of our results by constructing a new 3-charge gravity dual corresponding to a profile that breaks one of the two axial symmetries of $S^3$. In \S{7} we explicitly calculate the conserved charges for the special case geometry and show these results to be in agreement with the dual CFT. Finally, in \S{8} we summarize our results and discuss possible future work.

\newsection{Minimal Supergravity in Six Dimensions}\label{GMRsection}

In this section we begin with a brief discussion of minimal supergravity in six dimensions. The bosonic field content of the theory consists of a graviton $g_{\mu\nu}$ and a self-dual three-form. Solutions of this theory can be lifted to ten dimensional Type IIB supergravity by adding a flat $T^4$ and identifying the three-form as the field strength of the Ramond-Ramond 2-form gauge field. In a particular duality frame, solutions constructed in this manner carry equal D1 and D5 charges \cite{gm1}. Conversely, ten dimensional 2-charge solutions of Type IIB theory with the charges set equal  (so that the dilaton is trivial) and then reduced along the $T^{4}$ can be embedded in six dimensional supergravity.  

All supersymmetric solutions of this theory have been classified in~\cite{GMR}. In the following we review their construction and equations of motion.
\subsection{General Form of Supersymmetric Solutions in 6D}
In \cite{GMR} it was shown that all supersymmetric solutions of minimal 6D supergravity can be written in the following form (GMR form),
\bea
ds^2 &=& -2 H^{-1} (dv+\beta) \Bigl(du + \omega + {\F\over 2} (dv+\beta)\Bigr) + H h_{pq} dx^p dx^q\label{ansatz}\\
G^{(3)}&=&d_6 C^{(2)} \label{G} \\
&=& d_6\Bigl[
H^{-1}(dv+\beta)\wedge (du+\omega)\Bigr]+(dv+\beta)\wedge ({\mathcal{G}}^+ +2 \psi)+
\star_4 (\D H +H \dot{\beta}) \nonumber
\eea
The metric $h_{pq}$ defines the four-dimensional base space; $H$, $\F$ are 0-forms, $\omega$, $\beta$ are 1-forms while $\mathcal{G}^{+}$ and $\psi$ are 2-forms on the base space; $d_6$ is the exterior derivative on the six dimensional manifold. It is important to note that $H$, $\F$, $\beta$, $\omega$, and $h_{pq}$ will in general possess functional dependence on the $v$ coordinate. A derivative with respect to $v$ is denoted by a ``dot''. All the functions above are independent of $u$, hence $\frac{\partial}{\partial u}$ is a null Killing vector of the geometry as required by supersymmetry.

The base metric is also required to possess three almost complex structures satisfying the quaternionic algebra. In the following we will not distinguish the almost complex structures from the associated 2-forms. Let $J^i$ ($i=1,2,3$) be a basis of almost complex structures for the base metric $h_{pq}$: these are anti-self-dual 2-forms, with respect to $h_{pq}$, that satisfy 
\be
d J^i = \partial_v (\beta\wedge J^i) \label{cs}
\ee
Here and in the following, $d$ refers to the exterior derivative operator on the base space. One also defines an anti-self-dual 2-form $\psi$ which, loosely speaking, measures the $v$-dependence of the base metric,
\be
\psi = {H\over 16} \epsilon^{ijk} (J^i)^{pq} (\dot{J}^j)_{pq} J^k \label{psidef}
\ee 
The self-dual 2-form $\mathcal{G}^+$ is defined as
\be
\G = H^{-1} \Bigl((\D \omega)^+ + {\F\over 2} \D\beta\Bigr)\label{gplus}
\ee
where,  for any $p$-form $\Phi$, the operator $\D$ is given by 
\be
\mathcal D \Phi = d\Phi -\beta\wedge \dot{\Phi}
\ee
and
\be
(\D \omega)^+ ={\D\omega + \star_4 \D \omega\over 2}
\ee
(the operator $\star_4$ uses the metric $h_{pq}$). Finally introduce the 1-form
\be
L = \dot{\omega} + {\F\over 2}\dot{\beta} -{\D\F\over 2}
\ee
With the above definitions the equations of motions for metric (\ref{ansatz}) and the self-dual 3-form (\ref{G}) can now be written in terms of four-dimensional operators ($d,\star_4$) on the base space 
\bea
dJ^i &=& \partial_v (\beta\wedge J^i) \label{1}\\
\D \beta &=& \star_4 \D \beta \label{2} \\
\D\beta\wedge \G &=&-\D\Bigl(\star_4 (\D H + H \dot{\beta})\Bigr)
\label{3}\\
d(\G+2\psi) &=& \partial_v (\beta\wedge (\G+2\psi) + \star_4 (\D H + H \dot{\beta}))
\label{4}\\
\star_4\D(\star_4 L) &=& {H\over 2} h^{pq}\partial_v^2 (H h_{pq}) +{1\over 4}(\partial_v H h^{pq})\partial_v (H h_{pq})-2\dot{\beta}_pL^p\label{ee}\\  
&& -{1\over 2}  \star_4 \Bigl((\G+2\psi)\wedge (\G+2\psi)\Bigr)+2 H^{-1} \star_4 \Bigl((\D\omega)^- \wedge \psi\Bigr)
\nonumber
\eea
The first two equations (\ref{1}) and (\ref{2}) guarantee the existence of a Killing spinor so that the solution is supersymmetric. Equations (\ref{3}) and (\ref{4}) are the equations of motion for the field strength $G^{(3)}$, which is just the Bianchi identity $dG^{(3)}=0$ since $G^{(3)}$ is self-dual in six dimensions. Equation (\ref{ee}) is the only component of the Einstein equations that is not implied by  the previous equations.

\newsection{2-Charge Geometries}\label{2chargesection}
\subsection{Writing the D1-D5 Supergravity Solutions in GMR and GH Form}
We begin with the supergravity solutions dual to the set of Ramond ground states of the D1-D5 system~\cite{LM}. As discussed in the previous section we will take the D1 and D5 charges to be equal. In this case the non-trivial part of the geometry can be written in six dimensions as a two-dimensional fiber, spanned by coordinates $t$ and $y$, over a four-dimensional base space, which is just flat $\mathbb{R}^4$ \cite{gm1}. The metric and the 2-form gauge field are given by \cite{LM},
\bea
ds^2&=&(1+\bar{H})^{-1}\left[ -(dt-A)^2+(dy+B)^2 \right] + (1+\bar{H})\delta_{pq}dx^pdx^q \nonumber\\
&=&-2(1+\Hbar)^{-1} (dv+\bbar )(du+\obar )+(1+\Hbar) \delta_{pq}dx^pdx^q \label{2charge}\\\nonumber\\
C^{(2)}&=& (1+\bar{H})^{-1}\left[(dt-A)\wedge (dy+B) \right] +\bar C  \nonumber \\
&=& (1+\Hbar)^{-1} (dv+\bbar)\wedge (du+\obar) + \bar{C} \label{gauge}
\eea
where in order to write the metric in the GMR form we have introduced light-cone coordinates on the $(t,y)$-fiber and defined 1-forms, $\bbar$ and $\obar$, on the base space as
\bea
u={t+y\over\sqrt{2}},&\, &v={t-y\over\sqrt{2}}\\
\bbar=-\frac{A+B}{\sqrt{2}},&\,&\obar =-\frac{A-B}{\sqrt{2}}
\eea
The metric functions must satisfy the equations
\bea
d\bstar_4 d\Hbar = 0, &\, & d\bstar_4 d A= 0\label{Aeq}\\ \nonumber&\,&\\
d\bar C= {\bar\star}_4 d \Hbar, \label{ceq} &\, & dB={\bar\star}_4 d A\label{beq}
\eea
Note that these equations imply that the field strengths of $\bbar$ and $\obar$ are self-dual and anti-self dual respectively.

In \cite{LM} these geometries were constructed by performing a series of dualities on the geometry sourced by a fundamental string, multiply-wound on the $y$ circle and carrying momentum along $y$. By this construction the family of D1-D5 geometries was parameterized by a closed curve in $\mathbb{R}^4$, which in the FP picture represents the profile of the string carrying momentum.  With the profile function expressed as $x_i = F_i(v)\,,\, i=1,\ldots,4\,,\, v\in[0,L]$, the solutions of equations (\ref{Aeq}) can be written in terms of the profile as \cite{LM},
\bea
\bar{H}&=&{Q\over L} \int_0^{L}dv {1\over |\mathbf{x}-\mathbf{F}(v)|^2}\\
A&=&-{Q\over L} \int_0^{L}dv {\dot{F}_i\over |\mathbf{x}-\mathbf{F}(v)|^2} dx_i
\eea
The parameter $L$ is related to the length of the compactification circle of $y$ and $Q$ is related to the D5 charge. For details see~\cite{MathurRecentReview}. The forms $\bar C$ and $B$ are defined implicitly by equations (\ref{beq}). Furthermore, the profile function $\mathbf{F}(v)$ is constrained by 
\be
\dot{\mathbf{F}}^2=1
\ee
in order to embed the solutions in minimal supergravity.

\subsection{Computing $\bar C$}
The 2-form $\bar C$ appearing in the specification of the Lunin-Mathur geometries may be given an integral expression as follows. Recall that the defining equation for $\bar{C}$ is
\be
d\bar C = \bar *_4 d\bar{H}
\ee
where
\be
\bar{H}=\frac{Q}{L}\int_{0}^{L} \frac{dv}{|\mathbf{x}-\mathbf{F}(v)|^2}
\ee
The star operation above is to be performed with respect to the flat metric on $\field{R}^4$. If one sets $\mathbf{F}(v)=0$ in the expression for $H$, the dual is easily found to be
\be
\bar C= -\frac{Q}{2}\left( \frac{ x_1^2 +x_2^2- x_3^2-x_4^2}{x_1^2 +x_2^2+ x_3^2+x_4^2} \right)\frac{\sigma_{12} \wedge \sigma_{34}-  \sigma_{11} \wedge \sigma_{33}}{(x_1^2+x_2^2)(x_3^2+x_4^2)}\label{c2basic}
\ee
where
\be
\sigma_{11} = x_1 dx_1 + x_2 dx_2, \ \sigma_{12} =  x_1 dx_2 - x_2 dx_1, \  \sigma_{33} =  x_3 dx_3 + x_4 dx_4,\ \sigma_{34} =  x_3 dx_4 - x_4 dx_3
\ee 
The translational isometry of $\field{R}^4$ implies that the solution for a general $\mathbf{F}(v)$ is given by replacing $\mathbf{x}$ by $\mathbf{x}-\mathbf{F}(v)$ in equation (\ref{c2basic})  and integrating over $v$. 
\be
\bar C= \frac{Q}{2L}\int_{0}^{L}  \frac{dv}{|\mathbf{x}-\mathbf{F}|^2} \left[\frac{\sigma_2}{ (x_1-F_1)^2+ (x_2-F_2)^2} - \frac{\sigma_2}{ (x_3-F_3)^2+ (x_4-F_4)^2}  \right]\label{c2sol}
\ee
where
\bea
\sigma_2 &=& \tsigma_{12} \wedge \tsigma_{34} -\tsigma_{11}\wedge \tsigma_{33} \\
\tsigma_{11} &=& (x_1-F_1) dx_1 + (x_2-F_2) dx_2 \\ \tsigma_{12} &=&  (x_1-F_1) dx_2 - (x_2-F_2) dx_1 \\ \tsigma_{33} &=&  (x_3-F_3) dx_3 + (x_4-F_4) dx_4 \\  \tsigma_{34} &=&  (x_3-F_3) dx_4 - (x_4-F_4) dx_3
\eea
We will use this procedure in $\S{6}$ to derive an explicit example of a new 3-charge geometry with a single axial isometry.

\subsection{Base Space in Gibbons-Hawking Form}
It will be useful to write the base space $\mathbb{R}^4$ in Gibbons-Hawking (GH) form \cite{gibbonshawking} by changing from cartesian to GH coordinates as,
\bea
x_1 = 2\sqrt{r}\sin{\frac{\theta}{2}}\cos{\frac{\phi-\tau}{2}},&\ & x_2 = 2\sqrt{r}\sin{\frac{\theta}{2}}\sin{\frac{\phi-\tau}{2}} \\
x_3 = 2\sqrt{r}\cos{\frac{\theta}{2}}\cos{\frac{\phi +\tau}{2}}, &\ & x_4 = 2\sqrt{r}\cos{\frac{\theta}{2}}\sin{\frac{\phi +\tau}{2}} \nonumber
\eea
After applying the above coordinate transformation the base metric in (\ref{2charge}) becomes
\be
ds^2_4=\delta_{pq}dx^pdx^q=\bar{V}^{-1}(\ebar )^2+\Vbar ds^2_3
\ee
where $ds^2_3=dr^2+r^2d\theta^2+r^2\sin^2\theta d\phi^2$ is the flat metric on $\mathbb{R}^3$ and $\ebar$ denotes the einbein on the fiber
of GH space
\be
\ebar = d\tau+\chibar
\ee
$\Vbar$ and $\chibar$ are a 0-form and a 1-form respectively on $\mathbb{R}^3$ that satisfy the GH condition
\be
d_3 \Vbar = \star_3 d_3 \chibar
\ee
We denote by $d_3$ the exterior differential on $\mathbb{R}^3$ and by $\star_3$ the Hodge-star
operation with respect to $ds^2_3$. In our case of flat $\mathbb{R}^4$, we have explicitly
\be
\Vbar ={1\over r}\,,\quad \chibar = \cos\theta d\phi
\ee 
The Hodge star operation with respect to the base space $ds^2_4$ will be denoted as ${\bar\star}_4$, and we have chosen
the orientation\footnote{Note that in cartesian coordinates this orientation corresponds to $\epsilon_{1234}=-1$} such that $\epsilon_{\tau r \theta \phi}=+1$.

\subsection{Near-Horizon Limit}
The near-horizon limit of metric (\ref{2charge}) is obtained by dropping the one in $1+\bar{H}$, that is by sending $1+\bar{H}\rightarrow \bar{H}$.
Then, the 2-charge geometry in the near-horizon limit is given by,
\bea
ds^2_\mathrm{n.h.}&=&-2\Hbar^{-1} (dv+\bbar)(du+\obar)+\Hbar(\Vbar^{-1} (\ebar )^2 + \Vbar ds^2_3)\label{near}\\ \nonumber\\
C^{(2)}_\mathrm{n.h.}&=& \Hbar^{-1} (dv+\bbar)\wedge (du+\obar) + \bar{C} \label{cnh}
\eea

\newsection{Spectral Flow to 3-Charge System}\label{spectralflowsection}
Starting from the near-horizon geometry (\ref{near}), momentum can be added by a spectral flow
operation, which in the gravity picture is the following coordinate change \cite{bal,mm}
\be
\tau \rightarrow \tau + \gamma v \label{spectral}
\ee
where 
\be
\gamma =\frac{2\sqrt{2}n}{R_y}, \quad n\in \mathbb{Z}
\ee
$R_y$ is the radius of the $y$ circle. Note that, as the generic 2-charge geometry depends on $\tau$, the geometry after spectral flow depends on $v$, via the combination $\tau +\gamma v$. The action of spectral flow distinguishes $\tau$ from the other coordinates on $\field{R}^4$ and this is one of the primary reasons why it is more convenient to work in Gibbons-Hawking coordinates. This also motivates the following decomposition of the 1-forms $\bbar$ and $\obar$
\be
\bbar= \ebar\, \bbar_0 + \bbar_1 \,,\quad \obar =  \ebar\, \obar_0 + \obar_1 
\ee
where the 1-forms $\bbar_1$ and $\obar_1$ have components only along $\field{R}^3$. 
After applying the coordinate change (\ref{spectral}) to the metric (\ref{near}) the result can be brought back into the GMR form
\be
ds^2_\mathrm{n.h.}  = -2 \Htilde^{-1} (dv+\btilde) \Bigl(du + \otilde + {\Ftilde\over 2} (dv+\btilde)\Bigr) + \Htilde \htilde_{pq} dx^p dx^q
\label{nh}
\ee
with
\bea
\Htilde &=& {\Hbar\over 1+\gamma \bbar_0}\nonumber\\
{\Ftilde\over 2}&=& \gamma \obar_0 - \gamma^2 {\Hbar^2 \over 2 \Vbar (1+\gamma \bbar_0)}\label{tilde1}\\
\btilde &=& {\bbar\over 1+\gamma \bbar_0}\nonumber\\
\otilde &=& \obar - \gamma {\obar_0 \bbar\over 1+\gamma \bbar_0}-\gamma {\Hbar^2 \ebar\over \Vbar (1+\gamma \bbar_0)}+\gamma^2 {\Hbar^2 \bbar\over \Vbar (1+\gamma \bbar_0)^2}\nonumber
\eea
and
\bea
\htilde_{pq} dx^p dx^q &=& \Vtilde^{-1} (\etilde)^2 + \Vtilde ds^2_3\nonumber\\
\etilde &=& d\tau +\chitilde\,,\quad \Vtilde = \Vbar(1+\gamma \bbar_0)\,,\quad \chitilde = \chibar -\gamma \bbar_1
\label{tilde2}
\eea
Here and in the following all the quantities after spectral flow are to be evaluated at $\tau + \gamma v$ instead of $\tau$. Since spectral flow on the gravity side is simply a coordinate transformation, the new metric (\ref{nh}) by construction satisfies the GMR equations of motion and possesses the same $AdS_3\times S^3$ asymptotics and supersymmetries as the metric before the spectral flow.

\subsection{Notation}
Since we will need to make reference to several different metrics (i.e. 2-charge and 3-charge metrics with flat or $AdS$ asymptotics) there is potential for confusion regarding the domain on which operators or functions are defined. In an attempt to clarify matters we will make use of the following notational conventions:
\begin{itemize}
\item A bar indicates that the specified function or operator refers to the the near-horizon 2-charge metric (\ref{near}) prior to spectral flow. For example $\bar{\star}_4$ refers to the Hodge star operator with respect to the flat $\mathbb{R}^4$ base metric. 
\item A tilde refers to the metric (\ref{nh}), i.e. the 3-charge metric after spectral flow with $AdS_3\times S^3$ asymptotics. For example, $\tilde{\star}_4$ refers to the Hodge star operation with respect to the base metric (\ref{tilde2}). 
\item Neither a bar nor a tilde indicates that a function, form or operator is to be taken with respect to the 3-charge metric with flat asymptotics to be constructed in the next section.
\end{itemize}

\newsection{Extending to Flat Asymptotics}\label{add1}
Our aim is to extend the near-horizon geometry of (\ref{nh}) to obtain an asymptotically flat 3-charge solution. To this end, we should look for a solution where $\Htilde$ is replaced by, 
\be
H = 1+ \Htilde
\label{hchange}
\ee
In order to satisfy the equations of motion, we will also need to modify the other quantities appearing in the metric (\ref{nh}).
In general one has
\bea
\beta&=&\btilde+\delta\beta\\
\omega&=&\otilde+\delta\omega\\
h_{pq}&=&\htilde_{pq}+\delta h_{pq}\\
\F&=&\Ftilde +\delta \F
\eea
We need to find $(\delta\beta, \delta\omega, \delta\F,\delta h_{pq})$ such that the six dimensional metric defined by $H,\beta,\omega,$ $\F$ and $h_{pq}$ is asymptotically flat and satisfies the equations of motion given in (\ref{1})-(\ref{ee}). 

\subsection{Base Metric}
We know that by construction the near-horizon geometry satisfies the equations of motion. We make the 
ansatz that the base metric $h_{pq}$ of the asymptotically flat geometry equals the base metric $\htilde_{pq}$ of the near-horizon geometry
\be
h_{pq} dx^p dx^q = \htilde_{pq} dx^p dx^q \equiv   V^{-1} \ehat^2 + V ds^2_3\,,\quad \ehat = d\tau+\chi
\label{h}
\ee
with
\be
V=\Vtilde\,, \quad \chi=\chitilde
\ee
The main justification of this ansatz is that we know this is what happens in the axially-symmetric case \cite{Lunin}. The fact that the base metric remains invariant provides a great deal of simplifcation, for example the almost complex structures $J_i=\tilde J_i$ remain the same and the Hodge star operator is also invariant. Hence $*_4=\tilde{*}_{4}$. 

\subsection{$\beta$}
Under the assumption that the base metric is unchanged, the simplest way to satisfy equation (\ref{1}) is to assume that $\beta$ is also unchanged
\be
\beta=\btilde
\ee
Then equation (\ref{2}), the self-duality of the field strength of $\beta$, is automatically satisfied as well. 
The invariance of $\beta$ also greatly simplifies the process of solving for the remaining quantities. In particular it guarantees that the differential operator $\mathcal{D}$ remains invariant.

\subsection{$\psi$ and $\G$}
In order to find the new $\omega$ and $\F$, let us first look at the quantity $\G$ which is defined in terms of $(\D\omega)^+$ and $\F$ in (\ref{gplus}). First consider equation (\ref{3}): as $H$ changes as in (\ref{hchange}), it is clear that in order to satisfy this equation we must assume that $\G$ changes
\be
\G = {\tilde\mathcal{G}}^++\delta \G
\ee
Similarly, equation (\ref{4}) requires that both $\G$ and $\psi$ change. The change of $\psi$ can be derived directly from its definition (\ref{psidef}), given the fact that
$H$ changes as in (\ref{hchange}) and  the assumption that the base metric (and thus the complex structures $J^i$) do not change.  For a metric of the from (\ref{h}), a convenient choice of the
(almost) complex structures is 
\be
J^i = \ehat \wedge dx^i - V {\epsilon^{ijk}\over 2} dx^j\wedge dx^k
\ee
Substituting this into (\ref{psidef}) we find
\be
\psi = {H\over 4 V} \dot{\chi}_i J^i 
\ee
where
\be
\dot{\chi} = \gamma \partial_\tau \chi=\gamma \partial_\tau (\chibar -\gamma \bbar)= -\gamma^2 \partial_\tau \bbar
\ee
Similarly, for the near-horizon geometry one has 
\be
\tilde \psi = {\tilde H\over 4 V} \dot{\chi}_i J^i  
\ee
Thus, the change in $\psi$ is given by
\be
\delta \psi = \psi-\tilde\psi=-{\gamma^2\over 4 V}(\ehat \wedge \partial_\tau\bbar - V *_3 \partial_\tau \bbar)
\ee
Since $\G$ appears in the equations of motion in the combination $\G+2\psi$, one might  guess that the change in $\G$ must be related to $\delta\psi$;
more precisely, we take $\delta \G$ to be the ``self-dual'' version\footnote{For a metric of the form $ds^2_4=V^{-1}\ehat^2+Vds^2_3$, with a 1d fiber and 3d base space, a self-dual form on the full space can be constructed out of any 1-form $f_{3d}$ on the base space $ds^2_3$ as: $f_{(4d)}=\ehat\wedge f_{(3d)} + V*_3f_{(3d)}$. Anti-self dual forms are similarly given by $f_{(4d)}=\ehat\wedge f_{(3d)} - V*_3f_{(3d)}$} of $-2\delta \psi$
\be
\delta \G =  {\gamma^2\over 2 V}(\ehat \wedge \partial_\tau\bbar + V *_3 \partial_\tau \bbar)
\label{deltag}
\ee
One can now verify that the guess above solves both equations (\ref{3}) and (\ref{4}): using the fact that the ``tilded'' quantities satisfy the equations, it is sufficient to check that the
variations of (\ref{3}) and (\ref{4}) hold, i.e.
\bea
&&\D*_4 \dot{\beta}+\D\beta\wedge \delta \G=0\nonumber\\
&&d(\delta \G+2\delta \psi) = \partial_v (\beta\wedge (\delta \G+2\delta \psi) + *_4 \dot{\beta})
\eea

\subsection{$\F$ and $\omega$}
Having found the form of the new $\G$, one should determine $\omega$ and $\F$ in such a way
that equation (\ref{gplus}) is satisfied.\footnote{Eq. (\ref{gplus}) only involves $(\D \omega)^+$, and thus
determines $\omega$ only up to an ``anti-self-dual 1-form'', i.e. a 1-form $\delta\omega^-$ such that
 $(\D \delta \omega^-)^+=0$. In solving eq. (\ref{gplus}), we will have to guess the anti-self-dual part 
 of $\omega$. This guess will be justified by proving the validity of the Einstein equation (\ref{ee}), which
 explicitly involves $(\D \omega)^-$.} 
 One should have 
 \be
 (\D\delta\omega)^+ + {\delta\F\over 2}\D\beta = {\tilde\mathcal{G}}^+ + (\Htilde+1)\delta \G
 \label{deltaoeq}
 \ee
A relation involving ${\tilde\mathcal{G}^+}$, that will be useful in solving (\ref{deltaoeq}), can be found by looking at
the gauge field.  Start with the gauge field of the 2-charge system in the near-horizon region (\ref{cnh}) and apply the spectral flow transformation (\ref{spectral}). Writing the four dimensional  2-form $\bar C$
as
\be
\bar C=\ebar\wedge \bar C_1 +\bar  C_2
\ee
where $\bar C_1$ and $\bar C_2$ are a 1-form and 2-form along $\field{R}^3$, we find the gauge field of the spectral-flowed system to be
 \bea
 \tilde 
 C^{(2)}=\Htilde^{-1}(dv+\btilde)\wedge (du+\otilde)+\gamma(dv+\btilde)\wedge 
 \Bigl(\ehat {\Hbar\over V} + \bar C_1\Bigr) -\gamma\btilde \wedge \bar C_1 + \bar C
 \label{gauge1}
 \eea
 On the other hand, for a general supersymmetric solution the 3-form field strength should have
 the form given in equation (\ref{G})
 \bea
 \tilde G^{(3)}&=&d_6 \tilde C^{(2)}  \label{gauge2}\\
 &=& d_6 \Bigl[
\Htilde^{-1}(dv+\btilde)\wedge (du+\otilde)\Bigr]+(dv+\btilde)\wedge ({\tilde\mathcal{G}}^+ +2\tilde \psi)+
*_4 (\D \Htilde +\Htilde \dot{\btilde})\nonumber
 \eea
  Using the facts that the GMR ansatz has no dependence on $u$ and for our 3-charge metric dependence on $v$ is directly related to $\tau$ by (\ref{spectral}), the operator $d_6$ can be re-written as
 \be
 d_6=d+dv\wedge \partial_v = d+\gamma dv \wedge\partial_\tau
 \ee
 Define a 1-form
 \be
 \rho = -\gamma\Bigl({\Hbar\over V}\, \ehat  + \bar C_1\Bigr)
 \ee
 Comparison of (\ref{gauge1}) and (\ref{gauge2}) implies, after some lengthy algebraic manipulations, that
 \be
 \D \rho = {\tilde\mathcal{G}}^++ 2\tilde \psi+{\gamma^2\Htilde\over V}\ehat \wedge \partial_\tau \bbar +
 \gamma\Bigl(\ehat\wedge {\partial_\tau \bar C_1\over 1+\gamma \bbar_0}+\partial_\tau \bar C_2\Bigr)
 \label{abeq} 
 \ee
 If we take the self-dual part of (\ref{abeq}), and use (\ref{deltag}), we find
 \be
 (\D\rho)^+ =  {\tilde\mathcal{G}}^+ + \Htilde \delta\G  +{\gamma\over V}\Bigl(\ehat\wedge  \partial_\tau (\Vbar \bar C_1 +*_3\bar C_2) + V \partial_\tau ( \Vbar *_3 \bar C_1 + \bar C_2)\Bigr)
 \label{abeq2}
 \ee
 The last term on the r.h.s. of (\ref{abeq2}) can be made to vanish by taking the 2-form $\bar C$ to be
 anti-self dual, with respect to the flat base metric
  \be
  \bar C=-{\bar *}_4 \bar C\quad \Rightarrow\quad \Vbar *_3 \bar C_1 + \bar C_2=0
  \label{gaugec}
  \ee
Imposing condition (\ref{gaugec}) fixes the gauge invariance of the $\bar C$ defined by equation (\ref{ceq}). One can show that such a gauge can always be attained: for example, let $\bar C_L$ be the solution of  
(\ref{ceq}) in Lorentz gauge (which is certainly an allowed gauge),
\be
d{\bar *}_4 \bar C_L=0
\ee
Then 
\be
\bar C= \bar C_L-{\bar *}_4 \bar C_L
\ee
solves both (\ref{ceq}) and (\ref{gaugec}). \\\noindent
With this $\bar C$, equation (\ref{abeq2}) simplifies to
\be
(\D\rho)^+ =  {\tilde\mathcal{G}}^+ + \Htilde \delta\G \label{Drho+}
\ee
Now $\rho$ comes quite close to solving the equation for $\delta\omega$ (\ref{deltaoeq}). 
For a complete solution we need to add an additional piece in order to produce the extra $\delta \G$ term that is missing from (\ref{Drho+}). Let
\be
\delta\omega=\rho+\rho'
\ee
where the 1-form $\rho'$ should satisfy
\be
(\D \rho')^+= \delta \G - {\delta\F\over 2} \D\beta
\label{rhoprime}
\ee
We know from the solution of \cite{Lunin} that $\rho'$ and $\delta \F$ vanish in the axially symmetric (i.e. $\tau$-independent) case.  We have found only one solution of (\ref{rhoprime}) with this property: this solution has
\be
\delta \F=0
\ee
and a  $\rho'$ that can be constructed as follows. Decompose $\rho'$ as
\be
\rho'=\gamma^2(\ehat\, \rho_0'+\rho_1')
\ee
If we set $\rho_0'=0$, we find, using the expression for $\delta\G$ in equation (\ref{deltag}), that $\rho_1'$ must satisfy
\be
\Vbar \partial_\tau\rho_1'+*_3 (d_3-\bar\chi\partial_\tau )\rho_1' = \partial_\tau \bbar_1 \label{rho1eq}
\ee
In order to find a solution of (\ref{rho1eq}) we introduce a 2-form $\bar C'$ (analogous to the $\bar C$ used to construct $\rho$) on $\mathbb{R}^4$ 
satisfying
\be
d\bar C'= -{\bar*}_4\partial_\tau \bbar
\ee
Such a $\bar C'$ is guaranteed to exist if $\bbar$ is chosen to satisfy the Lorentz gauge
\be
d{\bar*}_4 \bbar =0
\ee 
Moreover take $\bar C'$ to be anti-self dual (with respect to flat metric on $\mathbb{R}^4$)
\be
\bar C'=-{\bar *}_4 \bar C'
\ee
As we explained above, this is a gauge choice for $\bar C'$ that can always be attained. Decompose $\bar C'$
as
\be
\bar C'=\ebar\wedge \bar C'_1 + \bar C'_2=\ebar\wedge\bar C'_1 -\Vbar *_3\bar C'_1
\ee 
Then a $\rho'$ satisfying  (\ref{rhoprime}) is given by
\be
\rho'=\gamma^2 \rho'_1 = \gamma^2 \bar C'_1
\ee
To summarize, we have found that
\be
\delta\omega = -\gamma\Bigl({\Hbar\over V}\,\ehat  + \bar C_1\Bigr)+\gamma^2 \bar C'_1\,,\quad \delta\F=0 \label{deltaomega}
\ee
We now have a complete ansatz for the asymptotically flat 3-charge geometry that by construction satisfies the first four equations of motion (\ref{1})-(\ref{4}). We are left to check that this ansatz satisfies the Einstein equation (\ref{ee}). Since the other four equations of motion only depend on the self-dual part of $\omega$, satisfying the Einstein equation provides a very non-trivial validation of the construction of the new metric functions. Using again the fact that the near-horizon geometry satisfies (\ref{ee}), it is sufficient to consider only the variation of (\ref{ee}), which is given by the
following equation
\bea
*_4\D(*_4 \partial_v \delta\omega) &\!\!=\!\!& {1\over 2} h^{pq}\partial_v^2 (\Htilde h_{pq})+ {\Htilde\over 2} h^{pq}\partial_v^2 h_{pq} +{1\over 2} h^{pq}\partial_v^2 h_{pq} \nonumber\\
&&+ 
{1\over 4}\partial_v h^{pq} \partial_v (\Htilde h_{pq})+ {1\over 4}\partial_v( \Htilde h^{pq})\partial_v h_{pq}+ 
{1\over 4}\partial_v h^{pq} \partial_v  h_{pq} -2\dot{\beta}_p(\partial_v \delta\omega)^p\nonumber\\  
&& -*_4 \Bigl(({\tilde \mathcal{G}}^++2{\tilde \psi})\wedge (\delta \G+2\delta \psi)\Bigr)
 -{1\over 2} *_4 \Bigl((\delta\G+2\delta\psi)\wedge (\delta\G+2\delta\psi)\Bigr)\nonumber\\
 &&+  2 \Htilde^{-1} *_4 \Bigl((\D\delta \omega)^- \wedge {\tilde \psi}\Bigr)
\label{3bis}
\eea
A straightforward but tedious computation shows that our ansatz satisfies this equation.

\subsection{Summary of the Asymptotically Flat Solution}
Starting from a 2-charge geometry specified by $\Vbar$, $\chibar$, $\Hbar$, $\bbar$ and $\obar$, spectral flow generates a 3-charge geometry with $AdS_3\times S^3$ asymptotics, described by $\Vtilde$, $\chitilde$, $\Htilde$, $\btilde$, $\otilde$, $\Ftilde$ given in equations (\ref{tilde1}), (\ref{tilde2}). The continuation of this geometry to the asymptotically flat region is then given by:
\bea
ds^2 &=& -2 H^{-1} (dv+\beta) \Bigl(du + \omega + {\F\over 2} (dv+\beta)\Bigr) + H\left( V^{-1}\hat{e}^2+Vds^2_3 \right) \label{summansatz}
\eea
where,
\bea
H&=&1+\Htilde\nonumber\\
V&=&\Vtilde\nonumber\\
\chi&=&\chitilde\label{summary}\\
\beta&=&\btilde\nonumber\\
\F &=& \Ftilde\nonumber\\
\omega&=& \otilde -\gamma\Bigl( {\Hbar\over V}\,\hat{e} +\bar C_1\Bigr)+\gamma^2 \bar C'_1\nonumber
\eea
Here $\bar C_1$ and $\bar C'_1$ are the 1-forms appearing in the decompositions
\be
\bar C=\ebar\wedge \bar C_1 + \bar C_2\,,\quad \,\,\bar C'=\ebar\wedge \bar C'_1 +\bar C'_2 \label{c1c1eq}
\ee
and $\bar C$ and $\bar C'$ are 2-forms on $\mathbb{R}^4$ satisfying
\bea
d\bar C&=&{\bar *}_4 d\Hbar\,,\quad \,\bar C\,=-{\bar *}_4 \bar C\label{c2c2eq}\\
d\bar C'&=&-{\bar *}_4\partial_\tau \bbar\,, \quad \bar C'=-{\bar  *}_4\bar C'\nonumber
\eea 
  
\newsection{Special Case: a new 3-charge solution}\label{example}
In order to provide a concrete example we have applied the general results of $\S{5}$ to a specific profile, which leads to a new asymptotically flat, $\tau$-dependent 3-charge geometry. For this geometry  we have also directly verified that the equations of motion are satisfied. \\\\
\noindent Consider the following profile function parameterized in cartesian coordinates,
\be
 \mathbf{F}(v) = \tilde{a}\left( \cos\frac{\lambda}{2} \cos \frac{2\pi v}{L},\,\cos\frac{\lambda}{2} \sin\frac{2\pi v}{L}, \,\sin\frac{\lambda}{2} \cos\frac{2\pi v}{L}, \,\sin\frac{\lambda}{2} \sin\frac{2\pi v}{L} \right)\label{profile1}
\ee
Note that this profile corresponds to an $SO(4)$ rotation of the axially-symmetric solution found in \cite{LM,bal,mm}. The matrix that accomplishes the necessary rotation can be written as, 
\be
\Lambda = \left(\begin{array}{cccc} \cos\frac{\lambda}{2} & 0 & -\sin\frac{\lambda}{2} & 0 \\
 0& \cos\frac{\lambda}{2} & 0 & -\sin\frac{\lambda}{2}  \\
  \sin\frac{\lambda}{2} & 0 & \cos\frac{\lambda}{2} & 0 \\
 0 & \sin\frac{\lambda}{2} & 0 & \cos\frac{\lambda}{2} \end{array} \right) \label{rot}
\ee
where the rotation is parametrized by $\lambda$. It is clear that for $\lambda=0$ this matrix becomes the identity and profile (\ref{profile1}) reduces to the one in~\cite{LM,bal,mm} which leads to an axially-symmetric metric. It is important to note that the spectral flow operation~(\ref{spectral}) singles out a particular coordinate of $\field{R}^4$, i.e.~$\tau$, and thus does not commute with $SO(4)$. Consequently our new 3-charge geometry cannot be produced by a rotation of the geometry of~\cite{Lunin,gms1}.

Due to the complexity of the metric functions describing the new 3-charge geometry, we will present its construction  in a series of steps. We start with the well know axially-symmetric 2-charge geometry found in \cite{LM,bal,mm} and then perform the rotation to produce the 2-charge geometry associated with (\ref{profile1}). We then write the 1-forms associated with this geometry in GMR form. Finally, we give the asymptotically flat 3-charge geometry using the results of $\S{4}$ and $\S{5}$.

\subsection{2-Charge Geometry}
We start with the axially symmetric 2-charge geometry defined by \cite{LM,bal, mm},
\be
ds^2=(H^{(0)})^{-1}\left[ -(dt-A^{(0)})^2+(dy+B^{(0)})^2 \right] + H^{(0)}\left(r(d\tau + \cos\theta 'd\phi ' )^2 + \frac{1}{r} ds^2_3\right) 
\ee
where
\bea
H^{(0)}&=&\frac{Q}{4r_c^{(0)}}\label{mm1}\\
A^{(0)}&=&\frac{Q}{8\sqrt{c}}\left(1-\frac{r+c}{r_c^{(0)}}\right) \left( d\phi'-d\tau' \right) \label{mm2}\\
B^{(0)}&=&\frac{Q}{8\sqrt{c}}\left(\frac{r-c}{r_c^{(0)}}-1 \right) \left(d\phi'+d\tau' \right)\label{mm3}
\eea
and 
\be
r_c^{(0)}=\sqrt{r^2+c^2+2rc\cos\theta'}\,,\quad c={\tilde{a}^2\over 4}
\ee
We must now apply the rotation matrix (\ref{rot}) to the above metric. It can be shown the rotation induces the following coordinate transformation in the GH frame
\bea
 \cos\theta' & = & \cos\theta \cos\lambda -\cos\tau \sin\theta \sin\lambda  \\
 \tan \tau' & = & \frac{\sin\tau \sin\theta }{\cos\lambda \cos\tau \sin\theta+\cos\theta \sin\lambda } \\
 \tan \phi' & = &  \frac{ \cos\lambda \sin\theta \sin\phi + \sin\lambda ( \cos\phi \sin\tau +\cos\theta \cos\tau \sin\phi )}{\cos\lambda \sin\theta \cos\phi - \sin\lambda(\sin\phi \sin\tau - \cos\theta \cos\tau \cos\phi )}
\eea
After applying the above coordinate transformation to equations (\ref{mm1})-(\ref{mm3}), we obtain the 2-charge geometry associated with the profile (\ref{profile1}). 
This rotated geometry is specified by the harmonic function $\bar H$ and the 1-forms $A$ and $B$,
\bea
\bar H&=&\frac{Q}{4r_c}\\
A&=& \frac{Q}{8 \sqrt{c} } \left(1-\frac{r+c}{r_c}\right) \left[ \frac{ \cos\theta -\cos\lambda}{1- \cos\theta'} d\tau - \frac{\sin\lambda\sin\tau}{1-\cos\theta'} d\theta + d\phi \right] \label{exA}\\
B&=& \frac{Q}{8 \sqrt{c} } \left(\frac{r-c}{r_c} -1\right) \left[ \frac{ \cos\theta +\cos\lambda}{1+ \cos\theta'} d\tau + \frac{\sin\lambda\sin\tau}{1+\cos\theta'} d\theta + d\phi \right] \label{exB} 
\eea
 where
\be
 r_c=\sqrt{r^2  + c^2 +2 r c \cos\theta'}
\ee
In the next section, where we give the asymptotically flat 3-charge solution, it will be more useful to have the above expressions in the GMR notation. This will allow us to directly apply the results of $\S{5}$.
From equations (\ref{exA}) and (\ref{exB}) we can write down the components of the 1-forms  $\bbar$ and $\obar$,
\bea
\bbar_0\equiv \bbar_\tau&=& -\frac{Q}{4r_c}\left[\frac{ \cos\lambda (r-r_c+ c\cos\theta' )+\cos\theta ((r_c-r)\cos\theta'-c) }{\sqrt{2c}(1-\cos^2\theta')}\right] \label{bbar0}\\
\bbar_r&=&0\\
\bbar_\theta&=&\frac{Q}{4r_c}\frac{ (r_c-r-c\cos\theta')\sin\lambda\sin\tau}{\sqrt{2c}(1-\cos^2\theta')} \\
\bbar_\phi&=& \frac{Q}{4r_c}\sqrt{\frac{c}{2}}
\eea
and,
\bea
\obar_\tau&=&-\frac{Q}{4r_c} \frac{\cos\lambda((r-r_c)\cos\theta'+c) + \cos\theta(r_c-r-c\cos\theta')}{\sqrt{2c}(1-\cos^2\theta')}\\ 
\obar_r&=&0\\
\obar_\theta&=&\frac{Q}{4r_c}\frac{((r_c-r)\cos\theta'- c)\sin\lambda\sin\tau}{\sqrt{2c}(1-\cos^2\theta')}   \\
\obar_\phi &=& \frac{Q}{4r_c}\frac{r-r_c}{\sqrt{2c}}\label{obarphi}
\eea

\subsection{New Asymptotically Flat 3-Charge Geometry}
The metric of the asymptotically flat 3-charge geometry has the standard GMR form and can be written as,
\bea
ds^2 &=& -2 H^{-1} (dv+\beta) \Bigl(du + \omega + {\F\over 2} (dv+\beta)\Bigr) + H\left( V^{-1}\hat{e}^2+Vds^2_3 \right) 
\eea
In this section we find expressions for the metric functions $\beta,\omega,\F,H,\chi,V$, in terms of the 2-charge fields $\bbar$ and $\obar$, which are given explicitly in equations (\ref{bbar0})-(\ref{obarphi}). 
From equations (\ref{tilde1}) and (\ref{summary}) one can find that $H$, $\beta$, and $\F$ are given by,
\bea
H&=&1+\frac{Q}{4 r_{c}\left(1+\gamma\bbar_0 \right)} \label{exH} \\\nonumber\\
\beta&=&\frac{\bbar}{\left(1+\gamma\bbar_0\right)}\label{exBeta}\\
\F&=&2\gamma\obar_\tau-\frac{r\gamma^2Q^2}{16r_c^2(1+\gamma\bbar_0)}\label{exF}
\eea
The base metric is completely specified by the three dimensional 1-form $\chi$ and the scalar function V, which for this case can be written as,
\bea
 V&=&\frac{1}{r}\left(1+\gamma\bbar_0\right)\\
 \chi&=& -\gamma\bbar_\theta d\theta-\gamma\bbar_\phi d\phi+\cos\theta\left(1+\gamma\bbar_0 \right)d\phi
\eea
From equation (\ref{summary}) $\omega$ is given by,
\be
\omega=\otilde -\gamma\Bigl(\ehat {\Hbar\over V} +\bar C_1\Bigr)+\gamma^2 \bar C'_1 + \omega_0
\ee 
Here $\omega_0$ denotes a constant 1-form, that is left undetermined by the equations of motion, and will be fixed
below by requiring asymptotic flatness. In order to find $\bar C_1$ and $\bar C_1'$ we must solve equations (\ref{c1c1eq}) and (\ref{c2c2eq}) which we will do in the following. The piece denoted by $\tilde \omega$ can be written in terms of $\bbar$ and $\obar$ by using eq. (\ref{tilde1})
\bea
\otilde&=&\obar - \frac{\gamma \obar_\tau }{1+\gamma\bbar_0}\bbar +\frac{\gamma r  Q^2}{16r_c^2(1+\gamma\bbar_0)^2} \left[\gamma\bbar -(1+\gamma\bbar_0)(d\tau+\cos\theta d\phi) \right] 
\eea
Asymptotic flatness requires that $\omega$ vanishes at large distances: we note that, for $r\gg 1$,   $\tilde \omega\rightarrow 0$, however the term proportional to $\ehat$ in $\omega$ behaves as, 
\be
{\bar H\over V}\,\ehat \to {Q\over 4} (d\tau + \cos\theta d\phi)
\ee
The term proportional to $d\tau$, in the above relation, can be cancelled by taking
\be
\omega_0 = - {Q\over 4} d\tau
\ee
The term proportional to $d\phi$, on the other hand, has to cancel against $\bar C_1$, so that we must require that
\be
\bar C_1 = O\Bigl({1\over r^2}\Bigr) dr + O\Bigl({1\over r}\Bigr) d\theta +\Bigl[ O\Bigl({1\over r}\Bigr)-{Q\over 4}\cos\theta\Bigr] d\phi
\label{asc1}
\ee
The 1-form $\bar C_1'$ must vanish at large distances as
\be
\bar C'_1 = O\Bigl({1\over r^2}\Bigr) dr + O\Bigl({1\over r}\Bigr) d\theta + O\Bigl({1\over r}\Bigr) d\phi
\label{asc1prime}
\ee

\subsubsection{Computation of $\bar C_1$}
From the analysis in $\S{5}$ we know that $\bar C_1$ defines an anti-self-dual 2-form,
\be
\bar C=\hat{\bar{e}}\wedge \bar C_1-\bar{V}\ast_3 \bar C_1 
\ee 
where $\bar C$ must satisfy
 \be
 d\bar C = \bstar_{4} d\bar H
 \label{ceqas}
 \ee
The solution for $\bar C$ may be found by using eq. (\ref{c2sol}). Though an exact computation of the integral is possible the result is quite messy. It will be sufficient for our purpose to find the asymptotic behavior of $\bar{C}$ using (\ref{c2sol}).  

\subsubsection{Asymptotic form of $\bar C_1$}
In the limit of large $r$ we find following expressions for the coordinate components of $\bar C_1$,
\bea
\bar C_{1,r}&=& \frac{cQ}{4r^2}   \sin\lambda\sin\tau\sin\theta  \\ 
\bar C_{1,\theta}&=& \frac{cQ}{4r}   \sin\lambda\sin\tau\cos\theta\\
\bar C_{1,\phi}&=& -\frac{Q}{4}\left( \cos\theta +\frac{c \sin\theta}{r}(\cos\tau\cos\theta\sin\lambda+\cos\lambda\sin\theta) \right) 
\eea
We note that the leading order behavior for the components satisfies the asymptotic boundary condition given in eq. (\ref{asc1}).

\subsubsection{Computation of $\bar C_1'$}
$\bar C'_1$ can be found by considering an anti-self-dual 2-form,
\be
\bar C' =  \hat{\bar{e}} \wedge \bar C_{1}' - \bar{V} *_3 \bar C_{1}' 
\ee
where $\bar C'$ must satisfy,
  \be
 d\bar C' =-\bstar_{4}  \partial_\tau \bbar
 \ee
For the specific geometry under consideration the solution to this equation is 
 \bea
 \bar C_{1}' &=& -\frac{Q}{4r_c}\frac{(r-r_c+c\cos\theta')\sin\lambda\sin\tau}{\sqrt{2c}(1-\cos^2\theta')} \left[ r \sin^2\theta d\theta -\sin\theta \cos\theta  dr \right] 
 \eea
 This $\bar C'_1$ vanishes at infinity as required by eq. (\ref{asc1prime}).

\newsection{Charges for the Special Case and Comparison with the CFT}
In this section we compute the asymptotic charges associated with the special case geometry presented in $\S{6}$. We also identify the corresponding state in the dual CFT and show that the field theory and gravity results are in agreement.

\subsection{Dimensional Reduction}
In order to compute the charges we must first dimensionally reduce to the five-dimensional Einstein frame. Starting from the general six-dimensional GMR ansatz (\ref{ansatz}) and completing the squares we have,
\bea
ds^2 &=& -2 H^{-1} (dv+\beta) \Bigl(du + \omega + {\F\over 2} (dv+\beta)\Bigr) + Hds_B^2\\
&=& -H^{-1}\left(1-\frac{\F}{2}\right)^{-1}\left( dt+ A^{(t)} \right)^2+H^{-1}\left(1-\frac{\F}{2}\right) \left(dy + A^{(y)}\right)^2+Hds_B^2 \nonumber
\eea
where 
\bea
A^{(t)}&=&\frac{\beta+\omega}{\sqrt{2}}\label{At}\\
A^{(y)}&=&(1-\frac{\F}{2})^{-1}\left[ \frac{\omega-\beta}{\sqrt{2}}+\frac{\F\beta}{\sqrt{2}}+\frac{\F}{2}dt \right]\label{Ay}
\eea
Now the five dimensional metric in the Einstein frame is
\be
ds^2_5=-\left(1-\frac{\F}{2}\right)^{-2/3}H^{-4/3}(dt+A^{(t)})^2+\left(1-\frac{\F}{2}\right)^{1/3}H^{2/3}ds_B^2\label{5d}
\ee
\subsection{Mass}
In the limit of large $r$ the $tt$ component of the metric is expected to behave as~\cite{myersperry}
\be
-g_{tt}\approx 1-{2 G^{(5)}\over 3 \pi}\frac{M}{r}
\ee
where $G^{(5)}$ is Newton's constant in 5D
\be
G^{(5)}={G^{(10)}\over V (2\pi R_y)} = {4 \pi^5 g^2\alpha'^4\over V R_y}
\ee
Here $V$ is the volume of $T^4$ and $g$ is the string coupling.
From metric (\ref{5d}) and using the explicit metric functions given in equations (\ref{exH}) and (\ref{exF}), we have the following expansion for large $r$
\be
-g_{tt}=(1-\frac{\F}{2})^{-2/3}H^{-4/3}\approx -1 + \frac{2Q + \tilde{a}^2n(n+\cos\lambda)}{6r}
\ee
By comparing the above expressions we can read off the mass of the geometry
\be
M={\pi\over 4 G^{(5)}}[2Q + \tilde{a}^2n(n+\cos\lambda)]
\ee

\subsection{Momentum}
In order to find the momentum charge consider the behaviour of the $t$ component of $A^{(y)}$.  In the limit of large $r$ this term behaves as \cite{myersperry}
\be
A^{(y)}_t\approx -\frac{Q_p}{4 r}
\ee
From equations (\ref{Ay}) and (\ref{exF})  we find
\be
A^{(y)}_t\approx \frac{\F}{2} \approx -\frac{ \tilde{a}^2 n(n+\cos\lambda)}{4r} 
\ee
This gives
\be
Q_p= \tilde{a}^2 n(n+\cos\lambda)
\ee
The quantized momentum charge $n_p$ is given by
\be
n_p = {\pi R_y \over 4 G^{(5)}} Q_p=  {\pi R_y \over 4 G^{(5)}} [ \tilde{a}^2 n(n+\cos\lambda)]
\ee
\subsection{Angular Momentum}
The angular momentum charges can be obtained from the five-dimensional $U(1)$ gauge field $A^{(t)}$, by considering \cite{Dabholkar},
\be
{1\over 2\pi R_y}\int^{2\pi R_y}_0 \!\!dy \,A^{(t)}= {1\over 2\pi R_y}\int^{2\pi R_y}_0\!\!dy\, \frac{\beta+\omega}{\sqrt{2}} \approx \frac{G^{(5)}}{\pi r}\left[ J_\psi \cos^2\theta'' d\psi''+J_\phi\sin^2\theta''d\phi'' \right]
\label{angular}
\ee
where the coordinates $r'',\theta'',\psi'',\phi''$ are the usual spherical polars for $\mathbb{R}^4$. These are related to the GH coordinates introduced in Section \ref{2chargesection} by
\be
r'' = 2\sqrt{r}, \quad \theta'' = \frac{\theta}{2}, \,\quad \psi'' = \frac{\tau+\phi}{2}, \,\quad \phi'' =\frac{-\tau+\phi}{2}
\ee
Then for the special case geometry given in $\S{6.2}$, we find that, 
\bea
J_\psi&=& {\tilde{a} \pi \over 16 G^{(5)}} [Q(1-2n-\cos\lambda)+2\tilde{a}^2 n (n+\cos\lambda)]
\label{wrong1}\\
J_\phi&=& {\tilde{a} \pi \over 16 G^{(5)}} [Q(1+2n+\cos\lambda)-2\tilde{a}^2 n(n+\cos\lambda)]\label{wrong2}
\eea
\subsection{Boost Needed}\label{boostsection}

The above results however lead to a problem because expressions (\ref{wrong1}) and (\ref{wrong2}) are not the charges expected from the CFT. An analogous problem was noted in~\cite{gms1}, where the axially-symmetric case was studied. In that case it was shown that extending the near-horizon geometry to the asymptotically flat region, using the $\tau$-independent analogue of the rules of $\S$5, produces a geometry in which the compact $y$ coordinate has the wrong periodic identification. By studying the regularity of the metric, it was also shown in~\cite{gms1} that the asymptotically flat metric can be made smooth by changing the periodic identifications on $y$; this can be formally achieved by performing a boost along $y$. This operation also changes the charges of the solution. 

In the case of our $\tau$-dependent geometry we expect a similar phenomenon to happen. The mismatch between the gravity and the CFT charges noted above signals the fact that a similar ``boost'' has to be performed in this case as well. A rigorous derivation of this statement would require a full singularity analysis which is beyond the scope of this paper. Assuming the necessity of a ``boost'' we show that it leads to a perfect agreement between the gravity and the CFT charges. The same boost is expected to desingularize the geometry. The coordinate transformation which produces the required ``boost'' is
\be
u\rightarrow u e^{-\sigma} ,\quad  v\rightarrow v e^{\sigma}\label{boost}
\ee
This change of coordinates acts on $\beta$, $\omega$ and $\F$ as
\be
\beta\to\beta e^{-\sigma}\,,\quad \omega\to\omega e^{\sigma}\,,\quad \F\to \F e^{2\sigma}
\ee
The momentum charge then transforms as
\be
Q_p\to Q_p e^{2\sigma} = \tilde{a}^2 e^{2\sigma} n (n+\cos\lambda)
\ee
If we also redefine the parameter $\tilde{a}$ as
\be
a=\ta e^{\sigma}
\ee
we see that the momentum charge after ``boost'' is
\be
Q_p = a^2  n(n+\cos\lambda)
\ee
The 1-form $A^{(t)}$ transforms as
\be
A^{(t)}\to {\beta e^{-\sigma} +\omega e^{\sigma} \over \sqrt{2}}
\ee
Expanding $A^{(t)}$ above according to (\ref{angular}) gives the
``boosted'' values of the angular momenta. It turns out that if we take 
\be
e^{-2\sigma}=\eta\equiv \frac{Q}{Q+2Q_p} \label{boostparam}
\ee
the ``boosted'' angular momenta assume the simple form 
\bea
J_\psi&=& {Q a \pi \over 16 G^{(5)}} (1-2n -\cos\lambda)\\
J_\phi&=& {Q a \pi \over 16 G^{(5)}} (1+2n +\cos\lambda)
\eea
We will see in the next subsection that these values match the ones expected from the CFT. It is important to note that the boost parameter in eq. (\ref{boostparam}) has the same form as the one found in~\cite{gms1} for the axially-symmetric case. Though the explicit value of $Q_p$ is different in the two cases, the boost parameter as a function of the charges of the solution is identical. This suggests that the eq. (\ref{boostparam}) gives the proper boost for the metric generated by an arbitrary profile.

In order to compare with the CFT it is convenient to rewrite the charges in terms of the
microscopic quantities. For this purpose we recall that the D1/D5 charge is quantized as
\be
Q={(2\pi)^4 g \alpha'^3 n_1\over V}={g \alpha' n_5}
\label{quant}
\ee
We also need the relation between the parameters  $a$, $Q$ and the 
radius $R_y$ of the compact direction $y$: based on consistency with the $\lambda=0$ case
\cite{gms1} we conjecture this relation to be
\be
R_y = {Q\over a}
\label{radius}
\ee
Note that the relation (\ref{radius}) and the boost described above, fix the global identifications of the geometry. These identifications are crucial to ensure that the metric is free of orbifold singularities.
Finally, using (\ref{quant}) and (\ref{radius}) we find
\bea
n_p &=& n_1 n_5 n (n+\cos\lambda)\\
J_\psi &=& {n_1 n_5\over 4}(1-2n-\cos\lambda)\, ,\quad J_\phi = {n_1 n_5\over 4}(1+2n+\cos\lambda)\label{angmo}
\eea

\subsection{Comparison with CFT}
The dual CFT is a (4,4) superconformal theory with central charge 
\be
c=6 n_1 n_5
\ee
It has an $SU(2)_L\times SU(2)_R$ R-symmetry that corresponds to $SO(4)$ rotations
on the 4 spatial non-compact directions of the gravity system. The diagonal generators
of  $SU(2)_L\times SU(2)_R$, $J_L$ and $J_R$, are connected to the space-time angular momenta
$J_\psi$ and $J_\phi$ as
\be
J_L=J_\phi - J_\psi\,,\quad J_R=J_\phi+J_\psi
\ee
The momentum charge $n_p$ is given by
\be
n_p = h_L-h_R
\ee
where $h_L$ and $h_R$ are the left and right conformal dimensions. The geometry we  
have constructed thus corresponds to a state with quantum numbers
\be
J_L = {n_1 n_5\over 2}(2 n + \cos\lambda)\,,\quad J_R = {n_1 n_5\over 2}\,,\quad h_L-h_R =
n_1 n_5 n (n+\cos\lambda)
\label{gravcharges}
\ee

We will show that these quantum numbers are the ones appropriate for a state obtained by 
applying spectral flow, in the left sector, to a Ramond ground state. Spectral flow is
an automorphism of the CFT superconformal algebra that acts independently on 
the left and right sectors and transforms the quantum numbers as
\bea
&&h_L\to h_L +\alpha_L J_L + \alpha^2_L {c\over 24}\,,\quad J_L\to J_L +\alpha_L {c\over 12}\nonumber\\
&&h_R\to h_R +\alpha_R J_R + \alpha^2_R {c\over 24}\,,\quad J_R\to J_R +\alpha_R {c\over 12}
\eea
If the parameters $\alpha_L$ and $\alpha_R$ are even integers, spectral flow sends the R sector to
itself and the NS sector to itself, if $\alpha_L$ and $\alpha_R$  are odd it interchanges the R and
NS sectors.

The 2-charge geometry (\ref{near}) is dual to a R ground state with
\be
J_L = {n_1 n_5\over 2} \cos\lambda\,,\quad J_R = {n_1 n_5\over 2}\,,\quad h_L=h_R=0
\ee
If we act on this state by spectral flow with parameters
\be
\alpha_L = 2n\,,\quad \alpha_R =0
\ee
we reach the R sector state with
\be
J_L = {n_1 n_5\over 2}(2 n + \cos\lambda)\,,\quad J_R = {n_1 n_5\over 2}\,,\quad h_L=
n_1 n_5 n (n+\cos\lambda)\,,\quad h_R=0
\ee
This agrees with the charges found in the gravity picture given in equation (\ref{gravcharges}).

\newsection{Discussion}

In summary, we have presented a new family of asymptotically flat, time-dependent, BPS solutions of Type IIB supergravity carrying D1, D5, and momentum charges. This set of solutions is parameterized by an integer $n$, defining the spectral flow of the left sector, and by a closed curve in $\field R^4$. These solutions represent the spectral flow of all Ramond sector ground states of the D1-D5 system. Constructing the solutions from known bound states of the D1-D5 system using an exact symmetry (spectral flow) guarantees that the new 3-charge solutions are also bound states of the branes. 
In addition to finding the general family of solutions we have also explicitly presented a new $v$-dependent 3-charge geometry. For this geometry we have calculated the conserved asymptotic charges and shown that the results are in agreement with the dual conformal field theory. 

A few outstanding issues pertaining to the new geometries presented here warrant further investigation. 
In the ``near-horizon limit'' our solutions are related to smooth geometries of~\cite{LM} by a globally defined coordinate transformation, and hence are regular in this limit. However, it is not \emph{a priori}  guaranteed that smoothness persists even after extending the geometries to the asymptotically flat region. Potential singularities may only appear at the origin of the base space and at
the location of the string profile. In the axially symmetric case it was shown \cite{gms1} that there exists, in both regions, a globally well defined coordinate transformation that brings the metric into a  manifestly smooth form. The smoothness of the geometry depends crucially on the global identifications imposed on the $y$ coordinate. It would be important to demonstrate that the periodic identifications on the $y$ coordinate introduced in  $\S$\ref{boostsection} lead to a smooth geometry even in our $\tau$-dependent case, though the task is expected to be significantly more involved. A more ambitious goal would be to show that the analogous identifications performed on the geometries corresponding to a general profile also lead to smooth solutions. The same analysis would shed some light on the qualitative features of the geometries.
 
Also, it is worth noting that the solutions presented here were constructed within the framework of six dimensional minimal supergravity, which requires that the D1 and D5 charges be equal ($Q_1=Q_5$). This condition made it possible to use the results of \cite{GMR} to find the correct asymptotically flat extensions of the near-horizon geometry after spectral flow. It would be interesting to find a generalized class of solutions without this restriction. 

Another useful extension of the calculations presented here would be to find expressions for the conserved charges corresponding to the general solution given in equations (\ref{summary})-(\ref{c2c2eq}) and to then compare the result with the dual conformal field theory. For this agreement to hold, we expect that the ``boost'' introduced in $\S$\ref{boostsection} would be crucial. Furthermore, a detailed investigation into the relationship between the CFT and this class of gravity duals could possibly lead to some new insight into the AdS/CFT duality and its relation to black hole microscopics. 

In conclusion, we have presented here a large class of 3-charge geometries, parameterized  
by a curve in $\mathbb{R}^4$, which generically break all but one (null) isometry of the 6D space, and 
have known CFT duals. We believe that these solutions represent a significant step towards the construction of general 3-charge microstates.

\section*{Acknowledgments}
The authors would like to thank Iosif Bena, Oleg Lunin, Samir Mathur, Amanda Peet, Simon Ross and Yogesh Srivastava, for many valuable discussions. S.G. and A.S. acknowledge the hospitality of the Aspen Center for
Physics during the initial stages of this work. J.F.  was supported by an Ontario Graduate Scholarship. S.G. and A.S. were supported by NSERC.

\bibliographystyle{amsalpha}

\end{document}